# Nonparametric Independence Screening via Favored Smoothing Bandwidth

Yang Feng

*Department of Statistics*
*Columbia University*
*New York, NY 10027*

Yichao Wu

*Department of Mathematics, Statistics, and Computer Science*
*University of Illinois at Chicago*
*Chicago, IL 60607-7045*

Leonard Stefanski

*Department of Statistics*
*North Carolina State University*
*Raleigh, NC, 27695-8203*

## Abstract

We propose a flexible nonparametric regression method for ultrahigh-dimensional data. As a first step, we propose a fast screening method based on the favored smoothing bandwidth of the marginal local constant regression. Then, an iterative procedure is developed to recover both the important covariates and the regression function. Theoretically, we prove that the favored smoothing bandwidth based screening possesses the model selection consistency property. Simulation studies as well as real data analysis show the competitive performance of the new procedure.

*Keywords:* bandwidth, nonparametric, smoothing, variable screening
*2010 MSC:* 62G08, 62H12

## 1. Introduction

High-dimensional data are increasingly available due to the advance of data collection and storage technology in assorted scientific fields such as biology, medicine, and finance. Such high-dimensional data provide many opportunities as well as challenges for statisticians. These challenges have motivated extensive research developed in the area of variable selection. In particular, the penalization framework for variable selection has been popularized by the lasso (Tibshirani, 1996). See Fan and Lv (2010) for a selective overview of penalization-based variable selection methods.

These penalization-based variable selection methods have shown to be very effective for variable selection. Yet their corresponding asymptotic properties typically hinge on stringent conditions. For

*Email address:* `yichaowu@uic.edu` (Yichao Wu)

example, Zhao and Yu (2006) established the sign (±) consistency for the lasso estimator under the irrepresentable condition. These conditions are challenging especially for the situation with an ultrahigh dimensionality, namely the dimensionality grows at an exponential rate of the sample size (Fan and Lv, 2008).

For ultrahigh-dimensional variable selection, Fan and Lv (2008) proposed sure independence screening for linear regression. Instead of fitting a huge joint model, the central idea of sure independence screening is to perform marginal regression for each predictor and rank them according to marginal utility. Then a refinement step is applied to the top ranked predictors, for example, by using a penalization-based method. In other words, they proposed a two-scale method for ultrahigh-dimensional variable selection by coupling a crude large-scale screening with a refined moderate-scale selection. They further established the sure screening property by proving that the screening retains important predictors with probability tending to one.

Since the introduction of sure independence screening, various extensions have been proposed for more general model settings. They include generalized linear model (Fan and Song, 2010), additive model (Fan et al., 2011), Cox model (Fan et al., 2010), varying-coefficient model (Liu et al., 2014; Fan et al., 2014) and binary classification (Mai and Zou, 2012) among many others. There has also been work on developing robust screening procedures (Zhu et al., 2011; Li et al., 2012a,b; Chang et al., 2013, 2016). All of these extensions have been proposed under some specific model assumption. Still lacking is a fully nonparametric ultrahigh-dimensional variable selection method. Our work fills this gap.

Stefanski et al. (2014) proposed a very general variable selection by intentionally adding measurement errors to predictors and devising a data-driven method to locate the best way to add measurement errors so that the loss of predictive power is minimized. It is named measurement error model selection likelihood. They proved that it reduces to the lasso when applied to linear regression. In this sense, it is an extension of lasso. They further illustrated it with nonparametric classification. In this case, it leads to a sparse version of the kernel discriminant analysis capable of performing variable selection for nonparametric classification. In an extension, White et al. (2015) studied the measurement error model selection likelihood in the context of nonparametric regression. It results in the Nadaraya–Watson estimator (Nadaraya, 1964; Watson, 1964) with variable selection capability. Their new method is called measurement error kernel regression operator.

A key observation from sparse kernel discriminant analysis and measurement error kernel regression operator is that an important predictor requires a small smoothing bandwidth while unimportant predictors favor an infinite smoothing bandwidth. This observation is echoed by Wu and Stefanski (2015). Wu and Stefanski (2015) focused on the additive model and used this key observation repeatedly to estimate the set of unimportant predictors, linear predictors, and so on towards polynomial modeling for additive models.

In this work, motivated by the above key observation that the smoothing bandwidth favored by each predictor is inversely connected with the predictor's importance in nonparametric regression, we propose a nonparametric screening method. The method is first to perform marginal nonparametric smoothing on each predictor and use an information criterion to determine its corresponding favored smoothing bandwidth. To improve the efficiency of the proposed screening method, for each predictor, we consider two candidate bandwidths and evaluate the corresponding information criterion values. The estimated signal set will be the variables that favors the smaller bandwidth. A two-scale fully nonparametric ultrahigh-dimensional variable selection is completed by applying the measurement error kernel regression operator to these top ranked predictors. Our idea of using favored smoothing bandwidth to rank predictors stands out quite uniquely in the literature of sure independence screening since most, if not all, of existing screening methods rank predictors according to correlation in one form or another.

To ensure the success of screening, we assume that if a variable $X_j$ is important, then the marginal relationship between $Y$ and $X_j$ is also strong in terms of the favored smoothing bandwidth. The details of this assumption will be described in Condition 7. A condition of such flavor is necessary for any screening method to succeed. Although it is by no means a very strong assumption, one could create scenarios to make it fail. In those scenarios, one possible remedy would be to apply a certain kind of iterative screening procedure (e.g., the IFBIS procedure to be introduced in Section 4.3) to identify the important variables sequentially.

The main contribution of this paper is twofold. First, a new nonparametric screening method is proposed based on the favored smoothing bandwidth for each predictor. It is shown that this favored bandwidth-based screening method possesses the model selection consistency property. Second, an iterative nonparametric variable selection and regression algorithm is developed that can handle different types of functional forms as well as interaction effects among covariates.

The rest of the paper is organized as follows. In Section 2, we introduce the nonparametric screening method via favored bandwidth. Its theoretical justification is provided in Section 3. Implementation issues are discussed and an iterative version is given in Section 4. Section 5 presents extensive simulation studies to illustrate its performance. Real data analysis are conducted in Section 6. We conclude with a short discussion in Section 7.

## 2. Method

Suppose $(\mathbf{X}_i, Y_i)$, $i = 1, \ldots, n$ are i.i.d. copies of $(\mathbf{X}, Y)$ with $\mathbf{X} = (X_1, \ldots, X_p)^T$ and $\mathbf{X}_i = (X_{i1}, \ldots, X_{ip})^T$, where $p$ denotes the number of predictors. We assume that the data are generated from the following very general nonparametric regression model

$$Y = g(\mathbf{X}) + \epsilon, \qquad (1)$$

where $E(\epsilon|\mathbf{X}) = 0$ and $\text{Var}(\epsilon|\mathbf{X} = \mathbf{x}) = \sigma^2(\mathbf{x})$. Denote $\text{Var}(Y) = \sigma_Y^2$.

Let $S \subset \{1, \ldots, p\}$ be the index set containing all the important covariates in predicting the response and denote $\mathbf{X}_S$ be the subvector of $\mathbf{X}$ with indices in the set $S$. We assume that $Y$ and $\mathbf{X}_{S^c}$ are independent given $\mathbf{X}_S$, where $S^c$ denotes the complement of $S$. In this case, $g(\mathbf{X})$ is a function of $\mathbf{X}_S$ only. More explicitly, there exists a function $g_S(\cdot)$ such that $g(\mathbf{X}) = g_S(\mathbf{X}_S)$. The goal is to recover the important variable set $S$ along with the estimation of the nonparametric regression function $g(\cdot)$.

Consider the scenario when $p$ is potentially much larger than the sample size $n$ (e.g., $p = o(\exp(n^a))$, for some $0 < a < 1$). For linear regression, Fan and Lv (2008) proposed the sure independence screening approach by examining each predictor individually and keeping the predictors with the largest marginal utility values. Here, we would like to develop a marginal screening procedure under the general model (1). Since we do not impose any parametric structure assumption on the regression function $g(\cdot)$, we consider the univariate nonparametric regression problem of regressing $Y$ on $X_j$ for each $j = 1, \ldots, p$.

Let $Y = g_j(X_j) + \epsilon_j$ with $g_j(X_j) = E(Y|X_j)$ and $\epsilon_j = g(\mathbf{X}) - g_j(X_j) + \epsilon$. When performing the univariate nonparametric regression, the function we are recovering is $g_j(\cdot)$, which can be thought of as a projection of $g(\cdot)$ onto the function space spanned by $X_j$. Note that $E(\epsilon_j|X_j) = 0$ by definition.

Consider now the conditional variance of $\epsilon_j$ given $X_j$. Denote $\sigma_j^2(x_j) = \text{Var}(\epsilon_j|X_j = x_j)$. If $j \in S$, we expect $g_j(\cdot)$ to be a nonconstant function and $\sigma_j^2(x_j)$ will be smaller than $\text{Var}(Y)$ on average because $X_j$ explains some of the variation of $Y$. On the other hand, if $j \notin S$, we expect $X_j$ to play little role in the marginal nonparametric smoothing, which implies that $g_j(\cdot)$ is a nearly constant function and consequently $E\sigma_j^2(X_j)$ is very close to $\sigma_Y^2$. Our approach exploits the differences among $\sigma_j^2(\cdot)$ for $j \in S$ and $j \notin S$ to identify the set $S$. The exact conditions on $\sigma_j^2(\cdot)$ for $j \in S$ and $j \notin S$ will be delineated in Section 3.

For the univariate nonparametric regression of $Y$ on $X_j$ based on data $\{(X_{ij}, Y_i), i = 1, \ldots, n\}$, we use the Nadaraya–Watson (NW) estimator defined as

$$\hat{g}_j(x) = \frac{\sum_{i=1}^n K_h(X_{ij} - x) Y_i}{\sum_{i=1}^n K_h(X_{ij} - x)}, \tag{2}$$

where $K_h(x) = h^{-1} K(x/h)$ for a kernel function $K(\cdot)$ and smoothing bandwidth $h > 0$.

Note that the NW estimator defined in (2) is a linear smoother as it linearly transforms the vector of responses $\mathbf{Y} = (Y_1, \ldots, Y_n)^T$ to the vector of fitted values $\hat{\mathbf{Y}}_j = (\hat{Y}_{1j}, \ldots, \hat{Y}_{nj})^T = (\hat{g}_j(X_{1j}), \ldots, \hat{g}_j(X_{nj}))^T$. The linear transformation can be represented in a matrix form as

$$\hat{\mathbf{Y}}_j = \mathbf{S}_{jh} \mathbf{Y},$$

where $\mathbf{S}_{jh} = \{S_{jh}(i, k), i = 1, \ldots, n; k = 1, \ldots, n\}$ with

$$S_{jh}(i, k) = \frac{K_h(X_{kj} - X_{ij})}{\sum_{l=1}^n K_h(X_{lj} - X_{ij})}. \tag{3}$$

Naturally, the residual sum of squares $RSS_j(h) = \sum_{i=1}^n \{Y_i - \hat{g}_j(X_{ij})\}^2$ measures the goodness of fit for the NW estimator $\hat{g}_j(\cdot)$ with smoothing bandwidth $h$. Except that the smoothing bandwidth is too small leading to overfitting, if $X_j$ is important, one would expect $RSS_j(h)$ to increase when $h$ increases because a larger smoothing bandwidth introduces bigger smoothing bias. It implies that important predictors favor a small smoothing bandwidth. On the other hand, if $X_j$ is unimportant, $RSS_j(h)$ will not change much when $h$ varies except a very small $h$ corresponding to overfitting. Its small change, if any, is due to stochastic variation. Correspondingly, an infinite smoothing bandwidth should be used for those unimportant predictors. Based on this intuition, to perform variable screening, a key step is to differentiate between predictors favoring a small smoothing bandwidth from those predictors favoring an infinite smoothing bandwidth.

To extract from data the information about which predictors favor a small smoothing bandwidth and which predictors favor an infinite smoothing bandwidth, we now introduce the following information criteria (IC) for each predictor variable

$$IC_j(h) = \log \left[ n^{-1} \sum_{i=1}^n \{Y_i - \hat{g}_j(X_{ij})\}^2 \right] + \tau [\text{tr}(\mathbf{S}_{jh}) - 1] (\frac{\log p}{n})^{1/2} h^{1/2}, \tag{4}$$

where $\tau > 0$ is a factor to control the penalty level, $\text{tr}(\cdot)$ represents the trace of a matrix and $\mathbf{S}_{jh}$ is the smoothing matrix defined in (3). The IC defined in (4) is similar to the form of AIC and BIC although the penalty term is specifically designed for the current nonparametric regression setting. This specific penalty helps to achieve the goal of differentiating between predictors favoring a small smoothing bandwidth from those predictors favoring an infinite smoothing bandwidth. The specific order of the penalty term represents the uniform estimation error across $p$ predictors, and the rationale will become clear after the theoretical properties are established in Section 3. We would like to comment that if one is considering the classical setting when $n \gg p$, it is necessary to replace the factor $\log p$ in (4) with $\log n$ to avoid under-penalization. Similar arguments go through for all the theoretical results to be presented.

When $h = \infty$, $\hat{g}_j(X_{ij}) = \bar{Y}$ and $\text{tr}(\mathbf{S}_{jh}) - 1 = 0$. As a result, we have

$$IC_j(\infty) = \log\left[n^{-1}\sum_{i=1}^{n}\{Y_i - \bar{Y}\}^2\right]. \tag{5}$$

For each predictor $X_j$, we can find its favored bandwidth according to $IC$ as follows

$$\hat{h}_j = \arg\min_h IC_j(h),$$

where the optimization is over any $h > 0$ not corresponding to extrapolation.

We can then use the favored smoothing bandwidth $\hat{h}_j$ as a measure of variable importance for the $j$-th predictor. The smaller $\hat{h}_j$ is, the more important the predictor $X_j$ is in the marginal nonparametric regression. Consequently, we can rank predictor variables according to their favored smoothing bandwidth $\hat{h}_j$ and keep those with small favored smoothing bandwidths to perform variable screening. However, searching over the entire space for the favored smoothing bandwidth $h$ as in (5) could be time consuming since we need to identify the favored smoothing bandwidth for all predictors. In practice, as a surrogate, we evaluate the value of $IC_j(h)$ at only two candidate bandwidths $h = h^* = (\frac{\log p}{n})^{1/5}$ and $h = \infty$ for each predictor. Here the candidate smoothing bandwidth $h^*$ is chosen at the optimal nonparametric rate. Then the estimated signal set is given by $\hat{S} = \{j | IC_j(h^*) < IC_j(\infty)\}$. Its theoretical properties will be studied in the next section. We name our method as favored bandwidth independence screening (FBIS).

## 3. Theory

In this section, we establish the theoretical properties of the favored bandwidth independence screening (FBIS) method. First of all, several technical conditions are introduced.

**Condition 1.** $\sup_u |K(u)| \leq \bar{K} < \infty$ and $\int |K(u)| du \leq \mu < \infty$.

**Condition 2.** *For some $0 < \Lambda_1 < \infty$ and $0 < L < \infty$, either $K(u) = 0$ for $|u| > L$ and for all $u$, $u' \in R$,*

$$|K(u) - K(u')| \leq \Lambda_1 |u - u'|,$$

*or $K(u)$ is differentiable, $|(\partial/\partial u)K(u)| \leq \Lambda_1$, and for some $\nu > 1$, $|(\partial/\partial u)K(u)| \leq \Lambda_1|u|^{-\nu}$ for $|u| > L$.*

Conditions 1 and 2 are standard conditions for kernels in kernel density estimation and Nadaraya-Watson estimates.

**Condition 3.** *Assume the number of predictors $p \geq n$ and $p = o(\exp(n^\alpha))$ with $0 < \alpha < 1$. For all $j = 1,\ldots,p$, $X_j$ has marginal density $f_j(x)$ with support $[0,1]$ such that $\sup_x f_j(x) \leq B_0 < \infty$ and $\sup_x E(|Y|^s | X_j = x) f_j(x) \leq B_1 < \infty$, for some $s > 2$. In addition, for $\delta_n = \inf_{j=1,\ldots,p} \inf_{x \in [0,1]} f_j(x) > 0$ and $h = o(1)$, we assume $\delta_n^{-1} a_n^* \to 0$ with $a_n^* = (\frac{\log p}{nh})^{1/2} + h^2$.*

The density assumption is necessary to ensure the proper behavior of the density estimates. A similar condition can be found in Hansen (2008). Note that here, we impose a uniform lower bound on the marginal density for each covariate, while allowing the bound decaying to 0 at a rate depending on $n$, $p$ and the bandwidth $h$.

For $i = 1,\ldots,n$ and $j = 1,\ldots,p$, define $U_{ij}(x) = Y_i K_h(x - X_{ij})$. For covariate $j$, we consider the average

$$\hat{\Psi}_j(x) = \frac{1}{n} \sum_{i=1}^n U_{ij}(x) = \frac{1}{nh} \sum_{i=1}^n Y_i K\left(\frac{x - X_{ij}}{h}\right).$$

**Lemma 1** (Bernstein's inequality, Lemma 2.2.11, Van Der Vaart and Wellner (1996)). *Let $Y_1,\ldots,Y_n$ be independent random variables with zero mean such that $E|Y_i|^m \leq m! M^{m-2} v_i / 2$, for every $m \geq 2$ (and all $i$) and some constants $M$ and $v_i$. Then*

$$\mathcal{P}(|Y_1 + \ldots + Y_n| > x) \leq 2 \exp\{-x^2 / (2(v + Mx))\},$$

*for $v \geq v_1 + \ldots v_n$.*

**Condition 4.** *Assume there exist constants $M$ and $v$, such that $E|U_{ij}(x) - EU_{ij}(x)|^m \leq m! M^{m-2} v / 2$ holds for all $i$, $j$, $h$, and any $x \in [0,1]$.*

This condition is imposed to facilitate the development of the uniform deviation results of $\hat{\Psi}_j$ towards its expectation, which is summarized in the following Theorem.

**Theorem 1.** *Assume Conditions 1-4 are satisfied. For $a_n = (\frac{\log p}{nh})^{1/2}$, we have the uniform deviation results as follows.*

$$\sup_{j=1,\ldots,p} \sup_{x \in [0,1]} |\hat{\Psi}_j(x) - E\hat{\Psi}_j(x)| = O_p(a_n).$$

*Proof.* The difficulty of the proof lies in the uniform bound across $p$ different covariates as well as the region of $x \in [0,1]$. In order to establish a uniform bound over $p$ covariates, we exploit the large deviation bounds for each covariate $X_j$. In particular, for a fixed $j$, we consider $|\hat{\Psi}_j(x) - E\hat{\Psi}_j(x)|$.

We consider intervals of the form $A_k = \{x : |x - x_k| \leq a_n h\}$. By selecting equal spaced $x_k$, the region $[0,1]$ can be covered with $N \leq (a_n h)^{-1}/2$ such regions like $A_k$. Condition 2 implies that for all $|x_1 - x_2| \leq \delta \leq L$, $|K(x_2) - K(x_1)| \leq \delta K^*(x_1)$, where $K^*(\cdot)$ is set as follows depending on the part of Condition 2 we assumes. If $K(u)$ has compact support and is Lipschitz, we set $K^*(u) = \Lambda_1 1\{|u| \leq 2L\}$. If $K(u)$ satisfies the differentiability condition with the bound on the derivative, we can set $K^*(u) = \Lambda_1 1\{|u| \leq 2L\} + |u - L|^{-\eta} 1\{|u| > 2L\}$. In both cases, $K^*(u)$ is bounded and integrable and therefore satisfies Condition 1. Note a similar argument can be found in the proof of Theorem 2 in Hansen (2008).

For any $x \in A_k$, $|x - x_k| \leq a_n h$, then from Condition 2, we have

$$|K(\frac{x - X_{ij}}{h}) - K(\frac{x_k - X_{ij}}{h})| \leq a_n K^*(\frac{x_k - X_{ij}}{h}).$$

Now, define

$$\tilde{\Psi}_j(x) = \frac{1}{nh} \sum_{i=1}^{n} Y_i K^*(\frac{x - X_{ij}}{h}),$$

which is a version of $\hat{\Psi}_j(u)$ with $K(u)$ replaced with $K^*(u)$. Note that by tower property of conditional expectation along with Condition 3, $E|\tilde{\Psi}_j(x)| \leq B_0 B_1 \int_R K^*(u) < \infty$. Then,

$$\sup_{x \in A_k} |\hat{\Psi}_j(x) - E\hat{\Psi}_j(x)| \leq |\hat{\Psi}_j(x_k) - E\hat{\Psi}_j(x_k)| + a_n[|\tilde{\Psi}_j(x_k)| + E|\tilde{\Psi}_j(x_k)|]$$

$$\leq |\hat{\Psi}_j(x_k) - E\hat{\Psi}_j(x_k)| + a_n|\tilde{\Psi}_j(x_k) - E\tilde{\Psi}_j(x_k)| + 2a_n E|\tilde{\Psi}_j(x_k)|$$

$$\leq |\hat{\Psi}_j(x_k) - E\hat{\Psi}_j(x_k)| + |\tilde{\Psi}_j(x_k) - E\tilde{\Psi}_j(x_k)| + 2a_n M_2,$$

for any $M_2 > E|\Psi_j(x_k)|$ since $a_n \leq 1$ for sufficiently large $n$.

As a result,

$$P(\sup_{x \in [0,1]} |\hat{\Psi}_j(x) - E\hat{\Psi}_j(x)| > 4M_2 a_n)$$

$$\leq N \max_{1 \leq k \leq N} \sup_{x \in A_k} |\hat{\Psi}_j(x) - E\hat{\Psi}_j(x)| > 4M_2 a_n)$$

$$\leq N \max_{1 \leq k \leq N} P(|\hat{\Psi}_j(x_k) - E\hat{\Psi}_j(x_k)| > M_2 a_n) + N \max_{1 \leq k \leq N} P(|\tilde{\Psi}_j(x_k) - E\tilde{\Psi}_j(x_k)| > M_2 a_n) \quad (6)$$

Recall $U_{ij}(x) = Y_i K_h(x - X_{ij})$ and define $U_{ij}^*(x) = Y_i K_h^*(x - X_{ij})$. Then, we have $\tilde{\Psi}_j(x) = n^{-1} \sum_{i=1}^{n} U_{ij}^*(x)$. Then, by Condition 4 and applying Bernstein's inequality in Lemma 1, we have for any $k = 1, \ldots, N$ and $n$ sufficiently large,

$$P(|\hat{\Psi}_j(x_k) - E\hat{\Psi}_j(x_k)| > M_2 a_n) = P(|\sum_{i=1}^{n} U_{ij}(x)| > M_2 a_n n)$$

$$\leq 2 \exp\left\{-\frac{a_n^2 n^2}{2(nv + Ma_n n)}\right\}$$

$$\leq 2 \exp\left\{-\frac{a_n^2 n}{2(v + Ma_n)}\right\}$$

$$\leq 2 \exp\left\{-\frac{\log p}{4hv}\right\}, \quad (7)$$

where we set $a_n = (\frac{\log p}{nh})^{1/2}$. Similarly, we have

$$P(|\tilde{\Psi}_j(x_k) - E\tilde{\Psi}_j(x_k)| > M_2 a_n) \leq 2 \exp\left\{-\frac{\log p}{4hv}\right\}. \quad (8)$$

Combining (6), (7) and (8), we have

$$P(\sup_{x \in [0,1]} |\hat{\Psi}_j(x) - E\hat{\Psi}_j(x)| > 4M_2 a_n) \leq 4N \exp\left\{-\frac{\log p}{4hv}\right\}$$

$$\leq 2(a_n h)^{-1} \exp\left\{-\frac{\log p}{4hv}\right\}$$

and

$$P(\sup_{j=1,\ldots,p} \sup_{x\in[0,1]} |\hat{\Psi}_j(x) - E\hat{\Psi}_j(x)| > 4M_2 a_n)$$

$$\leq 2p(a_n h)^{-1} \exp\left\{-\frac{\log p}{4hv}\right\}$$

$$\leq 2\exp\left\{-\frac{\log p}{4hv} + \log p - \log(a_n h)\right\}$$

Recall $a_n = (\frac{\log p}{nh})^{1/2}$, we have

$$P(\sup_{j=1,\ldots,p} \sup_{x\in[0,1]} |\hat{\Psi}_j(x) - E\hat{\Psi}_j(x)| > 3M_2 a_n) \to 0$$

as $n \to \infty$.

□

We now consider the NW estimator $\hat{g}_j(x)$ defined in (2). The aim is to create a uniform deviation results of $\hat{g}_j(x)$ towards its limit $g_j(x)$ over the range of $x$ and $j = 1, \ldots, p$. Before presenting the result, we need an additional condition adapted from Hansen (2008) to ensure that the NW estimators are well behaved.

**Condition 5.** *Assume the second order derivatives of $f_j(x)$ are uniformly continuous and bounded for $j = 1, \ldots, p$. The second order derivatives of $f_j(x)g_j(x)$ are uniformly continuous and bounded for $j \in S$.*

**Theorem 2.** *Assume Conditions 1-5 hold. For $a_n^* = (\frac{\log p}{nh})^{1/2} + h^2$, we have the uniform deviation results as follows,*

$$\sup_{j=1,\ldots,p} \sup_{x\in[0,1]} |\hat{g}_j(x) - g_j(x)| = O_p(a_n^*).$$

*Proof.* Set $g_j(x) = \Psi_j(x)/f_j(x)$, $\hat{\Psi}_j(x) = n^{-1}\sum_{i=1}^n Y_i K_h(x - X_{ij})$, and $\hat{f}_j(x) = n^{-1}\sum_{i=1}^n Y_i K_h(x - X_{ij})$. Then, we can write

$$\hat{g}_j(x) = \frac{\hat{\Psi}_j(x)}{\hat{f}_j(x)} = \frac{\hat{\Psi}_j(x)/f_j(x)}{\hat{f}_j(x)/f_j(x)}. \tag{9}$$

First, applying Theorem 1 by taking $Y_i \equiv 1$, we have

$$\sup_{j=1,\ldots,p} \sup_{x\in[0,1]} |\hat{f}_j(x) - f_j(x)| \leq O_p(a_n).$$

As a result,

$$\sup_{j=1,\ldots,p} \sup_{x\in[0,1]} \left|\frac{\hat{f}_j(x)}{f_j(x)} - 1\right| \leq \frac{O_p(a_n)}{\inf_{j=1,\ldots,p} \inf_{x\in[0,1]} f_j(x)} \leq O_p(\delta_n^{-1} a_n). \tag{10}$$

Again from the result of Theorem 1, we have

$$\sup_{j=1,\ldots,p} \sup_{x\in[0,1]} |\hat{\Psi}_j(x) - E\hat{\Psi}_j(x)| = O_p(a_n).$$

We now consider the uniform distance between $E\hat{\Psi}_j(x)$ and $\Psi_j(x)$ for each $j$ and any $x \in [0,1]$. It is easy to show

$$\begin{aligned}
E\hat{\Psi}_j(x) &= \frac{1}{h}E\left(E(Y_i|X_{ij})K(\frac{x-X_{ij}}{h})\right) \\
&= \frac{1}{h}\int K(\frac{x-u}{h})g_j(u)f_j(u)du \\
&= \int K(u)\Psi_j(x-hu)du \\
&= \Psi_j(x) + O(h^2)
\end{aligned}$$

Note that the $O(h^2)$ is uniform across all $j$ and $x \in [0,1]$ as we have a uniform bound for the second order derivative for $f_j g_j$ from Condition 5. Note that

$$\begin{aligned}
|\hat{\Psi}_j(x) - \Psi_j(x)| &\leq |\hat{\Psi}_j(x) - E\hat{\Psi}_j(x)| + |E\hat{\Psi}_j(x) - \Psi_j(x)| \\
&\leq O_p(a_n) + O(h^2).
\end{aligned}$$

As a result, for $a_n^* = a_n + h^2$, we have

$$\sup_{j=1,\ldots,p}\sup_{x\in[0,1]}|\hat{\Psi}_j(x) - \Psi_j(x)| = O_p(a_n^*). \tag{11}$$

By combining (10), (9) and (11), we arrived at

$$\sup_{j=1,\ldots,p}\sup_{x\in[0,1]}|\hat{g}_j(x) - g_j(x)| = O_p(\delta_n^{-1}a_n^*). \tag{12}$$

□

Theorem 2 shows the uniform consistency results for the Nadaraya-Waston estimator over $p$ predictors and the domain of $x$. The result itself is of interest. The theorem is an extension of the results in Hansen (2008). Note that to incorporate the growing dimensionality $p$, the uniform bound has a factor of $\log p$ instead of $\log n$.

Next, we study the uniform behavior of the $\log(\text{RSS}_j(h))$ when we regress $Y$ on $X_j$. The following moment condition is imposed.

**Condition 6.** Let $L_{ij} = \{Y_i - g_j(X_{ij})\}^2 - E\sigma_j^2(X)$ and $D_i = (Y_i - \bar{Y})^2 - \sigma_Y^2$. Assume $E|L_{ij}|^m \leq m!M^{m-2}v_i/2$ and $E|D_i|^m \leq m!M^{m-2}v_i/2$, for every $m \geq 2$ (and all $i$) and some constants $M$ and $v_i$.

This condition is necessary to establish the large probability uniform deviation bound (see equation (13) and (14) in the following proposition) of $\log(\text{RSS}_j(h))$ in the information criteria defined in (4). Note that here we use the same constant $M$ in both Conditions 4 and 6 to simplify presentation.

**Proposition 1** (Uniform Convergence). *Assume Conditions 1-6 are satisfied. For $j = 1,\ldots,p$, define*

$$L_j = n^{-1}\sum_{i=1}^n\{Y_i - g_j(X_{ij})\}^2, \widehat{L}_j = n^{-1}\sum_{i=1}^n\{Y_i - \hat{g}_j(X_{ij})\}^2,$$

where $\hat{g}_j$ is the Nadaraya-Watson estimate of $g_j$ with kernel $K$ and bandwidth $h$. There exists a set $\mathcal{A}_1$ with $P(\mathcal{A}_1) \to 1$ and a universal constant $A_1 > 0$ (does not depend on $n$ or $j$) such that on the set $\mathcal{A}_1$, for $j = 1, \ldots, p$,

$$|\log(\hat{L}_j) - \log(L_j)| \le A_1 \delta_n^{-1} a_n^*,$$

where $a_n^* = (\frac{\log p}{nh})^{1/2} + h^2$.

In addition, for $j = 1, \ldots, p$,

$$\mathcal{P}(|n^{-1}\sum_{i=1}^{n}\{Y_i - g_j(X_{ij})\}^2 - E\sigma_j^2(X)| > \delta) \le 2\exp\{-\frac{n\delta^2}{2[EL_{1j}^2 + M\delta]}\}. \tag{13}$$

Also,

$$\mathcal{P}(|n^{-1}\sum_{i=1}^{n}\{Y_i - \bar{Y}\}^2 - \sigma_Y^2| > \delta) \le 2\exp\{-\frac{n\delta^2}{2[ED_1^2 + M\delta]}\}. \tag{14}$$

In other words, by taking $\delta = A_2 n^{-1/2}$ and $A_3 n^{-1/2}$, there exist sets $\mathcal{A}_2$ and $\mathcal{A}_3$ with $P(\mathcal{A}_2) \to 1$ and $P(\mathcal{A}_3) \to 1$, such that, on the event $\mathcal{A}_2 \cap \mathcal{A}_3$, $|n^{-1}\sum_{i=1}^n\{Y_i - g_j(X_{ij})\}^2 - E\sigma_j^2(X)| \le A_2 n^{-1/2}$ and $|n^{-1}\sum_{i=1}^n\{Y_i - \bar{Y}\}^2 - \sigma_Y^2| \le A_3 n^{-1/2}$ for all $j = 1, \ldots, p$.

*Proof.* First, we observe that

$$|\log(\hat{L}_j) - \log(L_j)| \le \max(1/\hat{L}_j, 1/L_j)|\hat{L}_j - L_j|.$$
$$\le C_1 |\hat{L}_j - L_j|,$$

where $C_1$ is a universal constant that does not depend on $n$ or $j$.

From Theorem 2, there exists a set $\mathcal{A}_1$ with $P(\mathcal{A}_1) \to 1$ such that, on the set $\mathcal{A}_1$, for any $i = 1, \ldots, n$, $j = 1, \ldots, p$,

$$|\hat{g}_j(X_{ij}) - g_j(X_{ij})| \le \delta_n^{-1} a_n^*.$$

Then, we have

$$|\{Y_i - \hat{g}_j(X_{ij})\}^2 - \{Y_i - g_j(X_{ij})\}^2| \le |\hat{g}_j(X_{ij}) - g_j(X_{ij})||2Y_i - \hat{g}_j(X_{ij}) - g_j(X_{ij})|$$
$$\le C_2 \delta_n^{-1} a_n^*,$$

where $C_2$ is a uniform upper bound for $|Y_i - g_j(X_{ij})|$ over all $i$ and $j$ pairs. As a result, on the set $\mathcal{A}_1$, $|\hat{L}_j - L_j| \le C_2 \delta_n^{-1} a_n^*$ and $|\log(\hat{L}_j) - \log(L_j)| \le C_1 C_2 \delta_n^{-1} a_n^*$.

We now show (13). Note that we have $L_j - E\sigma_j^2(X) = n^{-1}\sum_{i=1}^n L_{ij}$, where $L_{ij} = \{Y_i - g_j(X_{ij})\}^2 - E\sigma_j^2(X)$. It is easy to see that

$$EL_{ij} = E[E[L_{ij}|X]] = 0.$$

Then, using Lemma 1 with Condition 6, we have,

$$\mathcal{P}(|\sum_{i=1}^{n} L_{ij}| > x) \le 2\exp\{-\frac{x^2}{2[nEL_{1j}^2 + Mx]}\}.$$

Choose $x = n\delta$ will lead to (13).

Following a similar argument for $D_i = (Y_i - \bar{Y})^2 - \sigma_Y^2$ with Condition 6, we have (14).

$\square$

**Condition 7.** *There exist sequences $C_n$ and $D_n$ such that $C_n \gg \delta_n^{-1}(\frac{\log p}{n})^{2/5} \geq D_n$ with $\delta_n$ specified in Condition 3. For all $j \in S$, $\sigma_Y^2 - E[\sigma_j^2(X)] > C_n$ and for all $j \notin S$, $\sigma_Y^2 - E[\sigma_j^2(X)] < D_n$. Here, for two sequences $a_n$ and $b_n$, we write $a_n \gg b_n$ to represent $b_n/a_n = o(1)$.*

The condition ensures that the signal level of the important covariates are detectable and puts an upper bound on the signal level of the unimportant variables. It resembles the usual beta-min condition imposed in the variable selection literature. Note that here we allow the signal level decaying with the sample size $n$ at certain rate and we do not assume the important covariates and unimportant covariates to be independent.

Before presenting the selection consistency results, we introduce the following Lemma for characterizing the uniform order of $tr(\mathbf{S}_{jh})$ over $j = 1, \ldots, p$. It is an extension of the corresponding results in Theorem 1 of Zhang (2003).

**Lemma 2.** *Assume Conditions 1, 3 and 5 are satisfied. We have*

$$\sup_{j=1,\ldots,p} h|tr(\mathbf{S}_{jh}) - \frac{\mathcal{K}(0)}{h}| \xrightarrow{p} 0,$$

*where $\mathcal{K}(0) = K(0)\mathbf{e}_1^T \mathbf{\Omega}^{-1}\mathbf{e}_1$ with $\mathbf{\Omega} = (\mu_{i+j-2})_{1 \leq i,j \leq p+1}$, in which $\mu_\ell = \int t^\ell K(t)dt$.*

*Proof.* From Theorem 1 of Zhang (2003), we know that the result of the theorem holds for each fixed $j$. Now, we argue that the result is true for the supreme of $p$ terms.

In the remaining of the proof, we use (Z.A.5) to represent the corresponding (A.5) in Zhang (2003) for simplicity. First, we examine (Z.A.5), it is clear that we can approximate the summation in (Z.A.2).

Here, we rewrite (Z.A.2) to emphasize the variable of consideration is $X_j$

$$A_{ij}(\ell, r) = (n-1)[E(X_j - X_{ij})^{\ell+r-2} K_h(X_j - X_{ij}) + o_p(1)],$$

where $o_p(1)$ is uniform across all $i$ and $j = 1, \ldots, p$. To prove uniformness of the small order, we can apply the Bernstein's inequality (Lemma 1) on the empirical average $\sum_{k=1}^n (X_{ik} - X_{ij})^{\ell+r-2} K_h(X_{ik} - X_{ij})$ and have a high probability bound on the deviation to the expectation. Note that the high-probability bound can be taken as a uniform bound for all $j$ due to Conditions 1 and 3. The detailed proof resembles that of Theorem 1 and is therefore omitted for brevity.

$$\begin{aligned}
&E(X_j - X_{ij})^{\ell+r-2} K_h(X_j - X_{ij}) \\
&= \int_x (x - X_{ij})^{\ell+r-2} K_h(x - X_{ij}) f_j(x) dx \\
&= \int_t (th)^{\ell+r-2} K(t) f(X_{ij} + th) dt \\
&= h^{\ell+r-2} \int_t t^{\ell+r-2} K(t)[f_j(X_{ij}) + th f_j'(X_{ij}) + 2^{-1} t^2 h^2 f_j''(X_{ij}) + o_p(h^2)] dt,
\end{aligned}$$

where the $o_p(h^2)$ is uniform across all $i$ and $j$, due to the uniform bound for the second order derivative of $f_j$ in Condition 5. As a result, (Z.A.5) holds uniformly for all $i$ and $j$.

We can use similar arguments for (Z.A.6)-(Z.A.9) to take care of the uniformness in $j$. Then, we have the desired uniform order specified in the Lemma for each variable $X_j$.

□

**Theorem 3.** *Assume Conditions 1-7 are satisfied. When $\tau \mathcal{K}(0) > (2A_1+1)\delta_n^{-1}$ with $\mathcal{K}(0)$ specified in Lemma 2, $A_1$ being the universal constant defined in Proposition 1 and $\tau$ being the penalty parameter in the information criteria* (4), *then, we have the selection consistency result for FBIS*

$$\mathcal{P}(\hat{S} = S) \to 1,$$

*as $n \to \infty$.*

*Proof.* To show $\hat{S} = S$ with high probability, we decompose the event into $p$ terms,

$$\begin{aligned}
\mathcal{P}(\hat{S} = S) &= \mathcal{P}\left(\cap_{j \in S}[IC_j(h^*) < IC_j(\infty)] \cap_{j \notin S}[IC_j(h^*) > IC_j(\infty)]\right) \\
&\geq 1 - \sum_{j \in S} \mathcal{P}(IC_j(h^*) > IC_j(\infty)) - \sum_{j \notin S} \mathcal{P}(IC_j(h^*) < IC_j(\infty)) \\
&\geq 1 - \sum_{j=1}^{p} \mathcal{P}(E_j).
\end{aligned}$$

where $E_j = \{IC_j(h^*) > IC_j(\infty)\}$ if $j \in S$ and $E_j = \{IC_j(h^*) < IC_j(\infty)\}$ if $j \notin S$. It remains to derive a lower bound of $\mathcal{P}(E_j)$ for each $j$.

Recall the definition for IC in (4). First of all, for $h = \infty$, we have

$$IC_j(\infty) = \log[n^{-1}\sum_{i=1}^{n}(Y_i - \bar{Y})^2]. \tag{15}$$

On the set $\mathcal{A}_2$ introduced in Proposition 1, $|\log[n^{-1}\sum_{i=1}^{n}(Y_i - \bar{Y})^2] - \log \sigma_Y^2| \leq A_2 n^{-1/2}$.

When $h = h^*$, we would expect the $IC_j(h)$ behave differently for $j \in S$ and $j \notin S$.

Now, consider the case when $h = h^* = (\frac{\log p}{n})^{1/5}$. We have

$$IC_j(h) = \log\left[n^{-1}\sum_{i=1}^{n}\{Y_i - \hat{g}_j(X_{ij})\}^2\right] + \tau[\text{tr}(\mathbf{S}_{jh}) - 1](\frac{\log p}{n})^{1/2}h^{1/2},$$

From Lemma 2, for any give constant $A_3 > 0$, there exists a set $\mathcal{A}_4$ with $P(\mathcal{A}_4) \to 1$ as $n \to \infty$ such that on the set $\mathcal{A}_4$, we have

$$|\text{tr}(\mathbf{S}_{jh}) - \frac{\mathcal{K}(0)}{h}| \leq \frac{A_3}{h},$$

for all $j = 1, \ldots, p$.

From Proposition 1, we have the uniform deviation of $\log \hat{L}_j$ and $\log L_j$ as follows. On the set $\mathcal{A}_1$,

$$|\log \hat{L}_j - \log L_j| \leq A_1 \delta_n^{-1} a_n^*.$$

When $h \to 0$ and $nh \to \infty$, for all $j$, we have on the set $\mathcal{A} = \mathcal{A}_1 \cap \mathcal{A}_2 \cap \mathcal{A}_3 \cap \mathcal{A}_4$,

$$\log E\sigma_j^2(X) + \tau\mathcal{K}(0)a_n - A_1\delta_n^{-1}a_n^* < IC_j(h) < \log E\sigma_j^2(X) + \tau\mathcal{K}(0)a_n + A_1\delta_n^{-1}a_n^*. \tag{16}$$

We are now ready to compare the $IC$ for the two choices of bandwidths regarding each variable. Assuming $h \to 0$ and $nh \to \infty$, for $j \in S$, we have on the set $\mathcal{A}$,

$$IC_j(h) - IC_j(\infty) < \log E\sigma_j^2(X) - \log \sigma_Y^2 + \tau\mathcal{K}(0)a_n + A_1\delta_n^{-1}a_n^* + A_2 n^{-1/2}.$$

With the choice of $h = (\frac{\log p}{n})^{1/5}$, and when $p = o(\exp(n^\alpha))$, we have

$$\tau \mathcal{K}(0) a_n - A_1 \delta_n^{-1} a_n^* + A_2 n^{-1/2} = \tau \mathcal{K}(0)(\frac{\log p}{n})^{2/5} + 2A_1 \delta_n^{-1}(\frac{\log p}{n})^{2/5} + A_2 n^{-1/2}$$

for sufficiently large $n$. Using Condition 7 on the signal level, we have for $j \in S$, $IC_j(h^*) - IC_j(\infty) < 0$ with high probability. For this reason, for the important variables, the favored bandwidth would be $h = h^*$.

Finally, we evaluate the IC values for $j \notin S$. Using (15) and (16), we have for $j \notin S$,

$$IC_j(h) - IC_j(\infty) > \log E\sigma_j^2(X) - \log \sigma_Y^2 + \tau \mathcal{K}(0) a_n - A_1 \delta_n^{-1} a_n^* - A_2 n^{-1/2}. \tag{17}$$

Now, for $h = h^* = (\frac{\log p}{n})^{1/5}$, for $j \notin S$, on the set $A$, if $\tau \mathcal{K}(0) > 2A_1 \delta_n^{-1}$, using Condition 7 for $j \notin S$, we have

$$IC_j(h^*) - IC_j(\infty) \geq (\tau \mathcal{K}(0) - (2A_1 + 1)\delta_n^{-1})(\frac{\log p}{n})^{2/5} - A_2 n^{-1/2} > 0.$$

As a result, for the unimportant variables, the favored bandwidth would be $h = \infty$.

$\square$

From Theorem 3, we observe that the proposed screening method can achieve the selection consistency under the ultrahigh-dimensional framework, i.e., $p = o(\exp(n^\alpha))$ with $0 < \alpha < 1$, in the sense that $P(\hat{S} = S) \to 1$ as $n \to \infty$. However, it is well known that using the one-step marginal screening method could miss variables that have weak marginal effects but are important given some other variables. To select those variables and improve finite sample performance, we need to develop an iterative version of the screening method, which is described in detail in the next section.

## 4. Implementation issues and iterative screening

### 4.1. An importance measure and vanilla screening

In Theorem 3, we showed that the proposed screening method is selection consistent under certain conditions. However, the proposed information criterion (4) has a super parameter $\tau$ that needs to be set properly. In practice, finding an appropriate choice of $\tau$ can be very challenging as its optimal choice depends on the unknown quantities as stated in Theorem 3. Consequently, it would be interesting to develop an importance measure that does not depend on $\tau$, which could then be used to generate a ranking for all the covariates.

Motivated by the information criterion defined in (4), we propose the following importance measure

$$IM_j = \frac{\log\left[n^{-1}\sum_{i=1}^n \{Y_i - \bar{Y}\}^2\right] - \log\left[n^{-1}\sum_{i=1}^n \{Y_i - \hat{g}_{j,h^*}(X_{ij})\}^2\right]}{\text{tr}(\mathbf{S}_{jh^*})(\frac{\log p}{n})^{1/2}(h^*)^{1/2}} \tag{18}$$

for each predictor $X_j$, where $\hat{g}_{j,h^*}(\cdot)$ is the corresponding NW estimator with bandwidth $h^* = (\log p/n)^{1/5}$. The importance measure has connection to the likelihood ratio test statistic. The numerator quantifies the change in terms of the residual sum of squares (therefore likelihood) between two choices of smoothing bandwidth. This change is then adjusted by taking into account the degrees of freedom. The predictor variables with a larger value of $IM$ would be regarded as more "important" in explaining the response.

After defining the above importance measure to be used for ranking predictors, the next issue is to choose an appropriate thresholding value. Fan and Lv (2008) suggested to keep the $\lfloor n/\log(n) \rfloor$ or $\lfloor n/(4\log(n)) \rfloor$ top ranked predictors. In practice, those choices may not work very well depending on the signal strength and sample size. Following Fan et al. (2011), we adopt a data-driven choice of thresholding value by permuting the sample. More explicitly, we generate a random permutation $\pi = (\pi(1), \ldots, \pi(n))$ of the indices $(1, \ldots, n)$. The random permutation is used to decouple $\mathbf{X}_i$ and $Y_i$ so that the resulting data $\{(\mathbf{X}_{\pi(i)}, Y_i), i = 1, \ldots, n\}$ follow a null model. Intuitively, after the random permutation, the corresponding importance measure would behave like that based on a random noise predictor variable. We calculate the $IM$ values for $\{(X_{\pi(i)j}, Y_i), i = 1, \ldots, n\}$ and denote them by $\widetilde{IM}_j, j = 1, \ldots, p$. Intuitively, the important predictor variables should have an $IM$ value larger than the majority of $\{\widetilde{IM}_j, j = 1, \ldots, p\}$. For a given quantile $q \in [0, 1)$, let $\omega_{(q)}$ be the $q$-th quantile of $\{\widetilde{IM}_j, j = 1, \ldots, p\}$. Then, our FBIS selects the following variables

$$\mathcal{A} = \{j : IM_j \geq \omega_{(q)}\}. \tag{19}$$

We refer to this step as the vanilla screening since it is based on marginal information only.

### 4.2. Refinement

After performing the above vanilla screening based on the favored smoothing bandwidth, we would like to use some more refined technique to fit the model with predictors in the estimated important set (19). As we do not impose any specific model assumption, a model-free technique would be highly desirable. White et al. (2015) proposed a nonparametric model selection method via measurement error selection likelihood (MEKRO). While their approach is flexible and works well in a wide range of settings, the computation cost is large for high-dimensional scenarios. Fortunately, our variable screening step has already reduced dimensionality to a moderate size, which could be well handled by MEKRO.

To be complete, we now provide details for the MEKRO corresponding to the set $\mathcal{A}$ selected above. Let $\mathbf{x}_{\mathcal{A}} = \{x_j, j \in \mathcal{A}\}$. For a kernel $K(\cdot)$ and smoothing bandwidth $h_j$ for each $j \in \mathcal{A}$, the multivariate Nadaraya–Watson estimator for the regression of $Y$ on $\mathbf{X}_{\mathcal{A}}$ based on data $\{(\mathbf{X}_{i\mathcal{A}}, Y_i), i = 1, 2, \ldots, n\}$ is given by

$$\hat{g}(\mathbf{x}_{\mathcal{A}}; \mathbf{h}_{\mathcal{A}}) = \frac{\sum_{i=1}^{n} Y_i \prod_{j \in \mathcal{A}} K_{h_j}(X_{ij} - x_j)}{\sum_{i=1}^{n} \prod_{j \in \mathcal{A}} K_{h_j}(X_{ij} - x_j)},$$

where $\mathbf{h}_{\mathcal{A}} = \{h_j, j \in \mathcal{A}\}$. Reparameterize $\lambda_j = 1/h_j$ for $j \in \mathcal{A}$ and define $\boldsymbol{\lambda}_{\mathcal{A}} = \{\lambda_j, j \in \mathcal{A}\}$ accordingly and $1/\boldsymbol{\lambda}_{\mathcal{A}} = \{1/\lambda_j, j \in \mathcal{A}\}$. Then the MEKRO achieves variable selection by solving the following optimization problem

$$\begin{aligned}
\text{minimize}_{\lambda_j, j \in \mathcal{A}} \quad & \sum_{i=1}^{n}(Y_i - \hat{g}(\mathbf{X}_{i\mathcal{A}}; 1/\boldsymbol{\lambda}_{\mathcal{A}}))^2 \\
\text{subject to} \quad & \lambda_j \geq 0, j \in \mathcal{A} \\
& \sum_{j \in \mathcal{A}} \lambda_j \leq \xi
\end{aligned}$$

for some regularization parameter $\xi > 0$. Denote the optimizer by $\hat{\boldsymbol{\lambda}} = \{\hat{\lambda}_j, j \in \mathcal{A}\}$. Then the refined estimator of the important set is given by $\mathcal{M}_1 = \{j : \hat{\lambda}_j > 0, j \in \mathcal{A}\}$.

*4.3. Iterative screening*

In linear regression, Fan and Lv (2008) demonstrated that the marginal screening may fail to retain predictors that are marginally unimportant but jointly important. Such predictors are possibly due to correlation. To deal with this issue, they proposed an iterative version and showed that the iterative screening can retain the aforementioned marginally unimportant but jointly important predictors very well. Next, we will propose an iterative version of our nonparametric screening.

Suppose the selected set is $\mathcal{M}_1$ as defined above after applying FBIS followed by MEKRO. We would like to create a conditional importance measure for each remaining predictor given those in $\mathcal{M}_1$. To achieve this, we first define $Z = \hat{Y}$ as a pseudo predictor, where $\hat{Y}$ is the fitted value of the nonparametric fit generated by MEKRO with the predictors in $\mathcal{M}_1$. The idea is that $Z$ contains most of the information of all the selected variables in $\mathcal{M}_1$. Then for each remaining candidate predictor $X_j, j \notin \mathcal{M}_1$, we consider the Nadaraya-Watson estimates by regressing $Y$ on $Z$ and $X_j$. In particular, we denote $\hat{g}_{j,h_1,h_2}(Z_i, X_{ij})$ as the fitted value using the bandwidths $h_1$ and $h_2$ for $Z$ and $X_j$, respectively. Then we define the conditional importance measure corresponding to $X_j$ given $\mathcal{M}_1$ as follows

$$IM_{j|\mathcal{M}_1} = \frac{\log\left[n^{-1}\sum_{i=1}^n\{Y_i - \hat{g}_{j,h^*,\infty}(Z_i, X_{ij})\}^2\right] - \log\left[n^{-1}\sum_{i=1}^n\{Y_i - \hat{g}_{j,h^*,h^*}(Z_i, X_{ij})\}^2\right]}{[\text{tr}(\mathbf{S}^j_{h^*,h^*}) - \text{tr}(\mathbf{S}^j_{h^*,\infty})](\frac{\log p}{n})^{1/2}(h^*)^{1/2}} \quad (20)$$

by mimicking the marginal unconditional importance measure defined in (18).

Note that in the definition of $IM_{j|\mathcal{M}_1}$ in (20), we compare two bivariate Nadaraya–Watson smoothing fits: one with smoothing bandwidths $h^*$ and $\infty$ for $Z$ and $X_j$, respectively while the other uses $h^*$ for both $Z$ and $X_j$, where $h^*$ is a small bandwidth. Since the pseudo predictor $Z$ is the surrogate of the selected important predictors in $\mathcal{M}_1$, it should be always treated as "important." That is why we use a small bandwidth for it in both bivariate smoothing fits used in the definition (20). The bivariate smoothing fit with smoothing bandwidths $h^*$ and $\infty$ is essentially a univariate smoothing fit with $Z$ only using smoothing bandwidth $h^*$, while the other bivariate smoothing fit with $h^*$ and $h^*$ corresponds to the fit with both $Z$ and $X_j$. In (20), the numerator compares the residual sum of squares corresponding to these two fits while the denominator adjusts the corresponding difference in terms of the degrees of freedom. Consequently $IM_{j|\mathcal{M}_1}$ measures how effective predictor $X_j$ is in reducing the residual sums of squares given $Z$, the surrogate of predictors selected in $\mathcal{M}_1$.

**Remark 1.** *Ideally, to measure the conditional importance of $X_j$ given predictors selected in $\mathcal{M}_1$, one should perform two $(|\mathcal{M}_1| + 1)$-dimension smoothings with $X_j$ and predictors in $\mathcal{M}_1$. In one smoothing, a small bandwidth is used for every predictor. The other smoothing uses a small bandwidth for all predictors in $\mathcal{M}_1$ and an infinity smoothing bandwidth for $X_j$. Then define the ratio similarly as in (20). However, it is well known that the nonparametric smoothing suffers the curse of dimensionality. That is why we introduce the surrogate $Z$ representing variables selected in $\mathcal{M}_1$ and use bivariate smoothing to define the conditional importance measure as in (20), which should serve as a good approximation. The benefit of using bivariate local constant smoothing is that we only need two smoothing bandwidths, compared to a general multivariate local constant smoothing with one smoothing bandwidth for each covariate whose computational load is prohibitive due to the curse of dimensionality. The bivariate local constant smoothing serves as a good approximation to the multivariate smoothing to trade off between bias and variance of the estimate and it is shown to work very well in numerical studies based on our limited numerical experience.*

Given data $\{(\mathbf{X}_i, Y_i), i = 1, \ldots, n\}$, the iterative favored bandwidth independence screening (IFBIS) method works as follows.

Step 1. Perform favored bandwidth independence screening using (19). Denote the selected set by $\mathcal{A}_1$.

Step 2. Apply MEKRO to the nonparametric regression of $Y$ on predictors in $\mathcal{A}_1$ with the tuning parameter selected by BIC. The selected set is called $\mathcal{M}_1$.

Step 3. Defined $Z = \hat{Y}$ as a surrogate predictor representing predictors selected in $\mathcal{M}_1$, where $\hat{Y}$ denotes the fitted value generated by MEKRO from Step 2.

Step 4. Apply bivariate local constant smoothing for the regression of $Y$ on $Z$ and $X_j$ for each $j \notin \mathcal{M}_1$. Calculate the conditional importance measure given $\mathcal{M}_1$ for each $j \notin \mathcal{M}_1$ using (20). Rank predictors $X_j$, $j \notin \mathcal{M}_1$ according to the conditional importance measure from the largest to the smallest and keep the top ranked ones. Denote the selected set by $\mathcal{A}_2$.

Step 5. Apply MEKRO to predictors in the set $\mathcal{M}_1 \cup \mathcal{A}_2$ with BIC tuning and the selected set is called $\mathcal{M}_2$.

Step 6. Iterate Steps 3-5 until $|\mathcal{M}_l| \geq s_0$ or $\mathcal{M}_l = \mathcal{M}_{l-1}$.

## 5. Numerical Studies

In this section, we evaluate the performance of the proposed Favored Bandwidth Independence Screening (FBIS) in terms of screening predictors and the iterative FBIS (IFBIS) in terms of both variable selection and regression function estimation. In particular, to demonstrate the screening performance of FBIS, we conduct comparison with SIS (Fan and Lv, 2008), NIS (Fan et al., 2011), DC-SIS (Li et al., 2012b) and SIRS (Zhu et al., 2011). For the iterative procedure IFBIS, we compare it with INIS (Fan et al., 2011).

Adapting the settings of Meier et al. (2009), Fan and Song (2010) and Fan et al. (2011), we consider the following numerical examples. For all examples, we fix $p = 1000$ and $n = 400$.

For simplicity of notations, we denote

$$g_1(x) = (2x - 1)^2, \quad g_2(x) = \frac{\sin(2\pi x)}{2 - \sin(2\pi x)}, \text{ and}$$

$$g_3(x) = 0.1\sin(2\pi x) + 0.2\cos(2\pi x) + 0.3\sin(2\pi x)^2 + 0.4\cos(2\pi x)^3 + 0.5\sin(2\pi x)^3.$$

**Example 1.** *Data are generated from the following additive model:*

$$Y = 4g_1(X_1) + 3g_2(X_2) + 3g_3(X_3) + \varepsilon$$

*with independent error $\varepsilon \sim N(0, \sigma^2)$.*

**Example 2.** *Data are generated from the following single-index model:*

$$Y = g_1(X_1 + X_2 - X_3 - X_4) + \varepsilon$$

*with independent error $\varepsilon \sim N(0, \sigma^2)$.*

**Example 3.** *Data are generated from the following model with interaction effects:*

$$Y = 4X_1 + 2sin(2\pi X_1)sin(2\pi X_2) + 3sin(2\pi X_2)sin(2\pi X_3) + \varepsilon$$

*with independent error $\varepsilon \sim N(0, \sigma^2)$.*

In all three examples, each predictor is marginally uniformly distributed over $[0, 1]$. The correlation among the $p$ uniformly distributed covariates is introduced via a monotonically transformed AR structure. In particular, we first generate multivariate Gaussian vectors $(\tilde{X}_1, \ldots, \tilde{X}_p)^T$ with mean $(0, 0, \ldots, 0)^T$ and covariance matrix $\Sigma$ satisfying $\Sigma_{jk} = \rho^{|j-k|}$ for all $j$ and $k$ pairs. Then, set $X_j = \Phi(\tilde{X}_j)$ for $j = 1, \ldots, p$, where $\Phi(\cdot)$ is the cumulative distribution function for the standard normal distribution. In all examples, we consider two correlation levels with $\rho = 0$ or $0.5$ as well as two different error variances $\sigma^2 = 1$ or $2$. This gives a total of four different combinations per example.

*5.1. Performance of vanilla screening*

For the screening performance of vanilla screening, we report in Table 1 the mean and standard error (in parentheses) of the number of selected important variables over 100 repetitions when we select the top 20 out of the total 1000 predictor variables. From Table 1, it is clear that in Example 2, all screening methods are able to capture all four important predictors. In Example 1, FBIS, DC-SIS and NIS perform better than SIS and SIRS. It is easy to understand why SIS is not performing competitively since it is a linear screening method. On first thought, it is surprising to observe that the model-free screening method SIRS does not have a good performance. Yet after checking the details, we found that the SIRS missed $X_1$ most of the time, especially for the independent predictor case with $\rho = 0$. The predictor $X_1$ has a symmetric quadratic effect, which is very challenging for sliced inverse regression (Li, 1991). The SIRS is based on the sliced inverse regression and consequently has difficulty in retaining $X_1$. For the correlated predictor case with $\rho = 0.5$, the correlation helps a lot in retaining $X_1$. In Example 3, none of these methods delivers a satisfying performance, especially for the case of independent predictors with $\rho = 0$. As a result, it would be necessary to conduct iterative screening to select all the important predictors.

Table 1: Average number of selected important variables over 100 repetitions with standard errors in corresponding parentheses.

| Ex $(\rho, \sigma^2)$ | FBIS | SIS | DC-SIS | SIRS | NIS |
|---|---|---|---|---|---|
| Ex 1(0,1) | 3.00(0.00) | 2.03(0.02) | 3.00(0.00) | 2.02(0.01) | 3.00(0.00) |
| Ex 1(0,2) | 3.00(0.00) | 2.03(0.02) | 3.00(0.00) | 2.02(0.01) | 3.00(0.00) |
| Ex 1(0.5,1) | 3.00(0.00) | 2.96(0.02) | 3.00(0.00) | 2.94(0.02) | 3.00(0.00) |
| Ex 1(0.5,2) | 3.00(0.00) | 2.95(0.02) | 3.00(0.00) | 2.94(0.02) | 3.00(0.00) |
| Ex 2(0,1) | 4.00(0.00) | 4.00(0.00) | 4.00(0.00) | 4.00(0.00) | 4.00(0.00) |
| Ex 2(0,2) | 4.00(0.00) | 4.00(0.00) | 4.00(0.00) | 4.00(0.00) | 4.00(0.00) |
| Ex 2(0.5,1) | 4.00(0.00) | 4.00(0.00) | 4.00(0.00) | 4.00(0.00) | 4.00(0.00) |
| Ex 2(0.5,2) | 4.00(0.00) | 4.00(0.00) | 4.00(0.00) | 4.00(0.00) | 4.00(0.00) |
| Ex 3(0,1) | 1.01(0.01) | 1.01(0.01) | 1.81(0.07) | 1.43(0.05) | 1.02(0.01) |
| Ex 3(0,2) | 1.00(0.00) | 1.01(0.01) | 1.33(0.05) | 1.21(0.04) | 1.02(0.01) |
| Ex 3(0.5,1) | 2.55(0.05) | 2.57(0.05) | 2.95(0.02) | 2.67(0.05) | 2.46(0.05) |
| Ex 3(0.5,2) | 2.41(0.05) | 2.41(0.05) | 2.74(0.04) | 2.47(0.05) | 2.37(0.05) |

*5.2. Performance of iterative screening*

Next we compare the newly proposed iterative procedure, IFBIS, with INIS in terms of three performance criteria: false positive (FP), false negative (FN) and mean squared prediction error (MSPE) evaluated on an independent test sample of size 10,000. Results over 100 repetitions are summarized in Table 2.

It shows that the IFBIS is able to identify all important variables in Example 1 with 0 false positive and a very small false negative (only 1 repetition with one important predictor missed out of 100 repetitions). It is worth noting that the IFBIS has a larger MSPE in Example 1 since the INIS benefits from the correct model specification by assuming the data is generated from an additive model, while the IFBIS makes no such assumption and the final fit is based on a multivariate local constant smoothing.

For Example 2, it is remarkable that the IFBIS achieves 0 false positive as well as 0 false negative across all 100 repetitions for all four different combination settings. Although both INIS and IFBIS are able to capture all the four important predictors, the INIS has a larger FP and a much bigger MSPE. The bigger MSPE is due to two sources. The first source is the large false positive predictors identified by the INIS. The other is due to the wrong additivity assumption used in the INIS since the true model is a single index model.

Example 3 is a model with complex interaction effects between $X_1$ and $X_2$ as well as $X_2$ and $X_3$ so that none of the vanilla independence screening methods works well in terms of picking up all three important predictors as reported in Table 1. By using the IFBIS, we are achieving perfect model selection consistency with both FP and FN equal 0. On the other hand, the INIS missed two important predictors on average when $\rho = 0$ and approximately 1 important predictor when $\rho = 0.5$. In addition, IFBIS leads to a much smaller MSPE than the INIS. In this example, in addition to the two sources as mentioned above for Examples 1 and 2, a third source, namely the big false negative predictors due to the INIS, contributes to the INIS's bad performance in terms of MSPE.

## 6. A real data example

In this section, we demonstrate the performance of the iterative favored bandwidth independence screening (IFBIS) on a real data set from Affymetrix GeneChip Rat Genome 230 2.0 Array. The data set was first presented in Scheetz et al. (2006) and later analyzed by Huang et al. (2010) and Fan et al. (2011). In this data set, the sample size is $n = 120$, representing twelve-week-old male rats selected for tissue harvesting from the eyes. The microarrays used to analyze the RNA from the eyes of these animals contain over 31,042 different probe sets (Affymetrix GeneChip Rat Genome 230 2.0 Array). The intensity values were normalized using the robust multi-chip averaging method (Irizarry et al., 2003) to obtain summary expression values for each probe set. Gene expression levels were analyzed on a logarithmic scale.

Following Huang et al. (2010) and Fan et al. (2011), we would like to identify the genes that are related to the gene TRIM32, which was found to cause Bardet-Biedl syndrome (Chiang et al., 2006), a genetically heterogeneous disease of multiple organ systems including the retina. Although over 30,000 probe sets are represented on the Rat Genome 230 2.0 Array, many of them are not expressed in the eye tissue. Thus, we focus on the 18975 probes which are expressed in the eye tissue. Following Huang et al. (2010), we first standardized each probe to have mean 0 and variance 1, then use 1000 probe sets that are expressed in the eye and have highest absolute marginal correlation with TRIM32 in the analysis. On the subset of the data ($n = 120, p = 1000$), we apply the IFBIS and INIS to model the relationship between the expression of TRIM32 and those of the 1000 probes.

Table 2: ($n = 400$) Average false positive (FP), false negative (FN) and mean squared prediction error (MSPE) of IBIS and INIS over 100 repetitions. Standard errors are in parentheses.

| Ex $(\rho, \sigma^2)$ | INIS | | | IFBIS | | |
|---|---|---|---|---|---|---|
| | FP | FN | MSPE | FP | FN | MSPE |
| Ex 1(0,1) | 2.06(0.21) | 0.00(0.00) | 1.40(0.01) | 0.00(0.00) | 0.00(0.00) | 2.02(0.02) |
| Ex 1(0,2) | 2.12(0.20) | 0.00(0.00) | 2.52(0.03) | 0.00(0.00) | 0.01(0.01) | 3.32(0.04) |
| Ex 1(0.5,1) | 2.79(0.27) | 0.00(0.00) | 1.41(0.01) | 0.00(0.00) | 0.00(0.00) | 1.98(0.02) |
| Ex 1(0.5,2) | 2.94(0.29) | 0.00(0.00) | 2.55(0.03) | 0.00(0.00) | 0.00(0.00) | 3.26(0.03) |
| Ex 2(0,1) | 1.97(0.22) | 0.00(0.00) | 4.09(0.05) | 0.00(0.00) | 0.00(0.00) | 1.98(0.03) |
| Ex 2(0,2) | 2.03(0.19) | 0.00(0.00) | 5.22(0.06) | 0.00(0.00) | 0.00(0.00) | 3.25(0.04) |
| Ex 2(0.5,1) | 3.03(0.23) | 0.00(0.00) | 4.01(0.05) | 0.00(0.00) | 0.00(0.00) | 1.87(0.03) |
| Ex 2(0.5,2) | 2.77(0.23) | 0.00(0.00) | 5.14(0.06) | 0.00(0.00) | 0.00(0.00) | 3.12(0.04) |
| Ex 3(0,1) | 2.48(0.20) | 2.00(0.00) | 4.57(0.04) | 0.00(0.00) | 0.00(0.00) | 1.83(0.02) |
| Ex 3(0,2) | 2.62(0.21) | 2.00(0.00) | 5.67(0.05) | 0.00(0.00) | 0.00(0.00) | 3.06(0.03) |
| Ex 3(0.5,1) | 2.21(0.22) | 0.85(0.04) | 4.43(0.04) | 0.00(0.00) | 0.00(0.00) | 1.80(0.02) |
| Ex 3(0.5,2) | 2.44(0.24) | 0.93(0.04) | 5.57(0.06) | 0.00(0.00) | 0.00(0.00) | 3.05(0.03) |

Table 3: Median model size (MS) and prediction mean squared error (PMSE) over 100 repetitions and their robust standard deviations(in parentheses) for IFBIS and INIS.

| Method | MS | PE |
|---|---|---|
| INIS | 8.00(0.75) | 0.384(0.211) |
| IFBIS | 4.00(0.75) | 0.377(0.248) |

To evaluate the performances of the two methods, we first randomly partition the data into a training set of 110 observations and a test set of 10 observations. Then we apply the method on the training data and compare the prediction mean squared error (PMSE) on the test data. During the process, we also record the number of probes selected by the two methods. This process is repeated 100 times. Table 3 presents the median values and their associated robust estimates of the standard deviation (RSD=IQR/1.34) over 100 replications. It is clear in the table that by applying the IFBIS approach, the number of probes selected is around half of the number selected when the INIS method is applied. In addition, the IFBIS approach leads to a slightly smaller median prediction error. One potential explanation of the result is that there may exist certain complicated functional regression relationship, like interaction effects, among the few selected probes that lead to a better prediction compared with an additive model.

## 7. Discussion

In this work, we propose a flexible nonparametric screening and selection method which is shown to work well in a wide range of settings. Here, we assume the smoothness of each marginal predictor are of the same order, which may not be the case in practice. How to extend the current results to the case where each predictor can have its own smoothness level will be an interesting future work. Another future research topic is to extend the methodology and the associated theory to the case of classification and categorical response.

### Acknowledgement

The authors thank the entire review team for suggestions and comments that lead to substantial improvements in the paper. Y. Feng was funded by NSF CAREER grant DMS-1554804; Y. Wu by NSF grant DMS-1055210 and NIH grant P01CA142538; and L. Stefanski by NSF grant DMS-1406456, NIH grants R01CA085848 and P01CA142538.